\documentclass[journal,comsoc, 10pt]{IEEEtran}

\usepackage{epsfig}
\usepackage{cite}
\usepackage{amsmath,amssymb,amsfonts}
\usepackage{textcomp}
\usepackage[usenames]{color}
\usepackage{psfrag}
\usepackage{arydshln}
\usepackage{url}
\usepackage{soul}
\usepackage{comment}
\usepackage{hyperref}
\usepackage{graphicx,color}
\usepackage{epstopdf}
\usepackage[nolist]{acronym}
\usepackage{url}
\usepackage[capitalise]{cleveref}
\Crefname{equation}{Eq.\!}{Eqs.\!}
\Crefname{figure}{Fig.\!}{Figs.\!}
\Crefname{tabular}{Tab.\!}{Tabs.\!}
\Crefname{section}{Section\!}{Sections.\!}

\usepackage[linesnumbered,ruled,vlined]{algorithm2e}
\usepackage{algorithmicx}
\usepackage{algpseudocode}
\usepackage{pgfplots}
\usepackage{pgfplotstable}

\newcommand{\CASE}[1]{\STATE \textbf{case} #1\textbf{:} \begin{ALC@g}}
	\newcommand{\ENDCASE}{\end{ALC@g}}

\newcommand{\DEFAULT}{\STATE \textbf{default:} \begin{ALC@g}}
	\newcommand{\ENDDEFAULT}{\end{ALC@g}}
\newcommand{\DEFAULTLINE}[1]{\STATE \textbf{default:} }

\begin{document}
	\newcommand{\norm}[1]{\left\lVert#1\right\rVert}
\newcommand{\normo}[1]{{\left\lVert#1\right\rVert}_{0}}
\def\SAR{\textrm{SAR}}
\def\Thetac{\boldsymbol{\Theta}}

\def\Hone{\mathbf{H}^{\mathrm{u}}}
\def\Htwo{\mathbf{H}^{\mathrm{r}}}
\def\Hthree{\mathbf{H}^{\mathrm{d}}}

\def\Honebar{\overline{{{\mathbf{H}}}}^{\mathrm{u}}}
\def\Hthreebar{\overline{{{\mathbf{H}}}}^{\mathrm{d}}}

\def\q{q_{m,k}}

\def\phin{\phi_{n,n}}
\def\x{\mathbf{x}}
\def\n{\mathbf{n}}
\def\u{\mathbf{u}}
\def\EI{\mathrm{EI}}

\def\minus{-p_{\mathrm{max}} \|\boldsymbol{\lambda}\|_1}
\def\Q{\mathbf{Q}}
\def\T{\boldsymbol{\mathrm{T}}}
\def\R{\boldsymbol{\mathrm{R}}}



\def\sign{\textrm{sign}}                                              
\def\erf{\textrm{erf}}
\def\erfc{\textrm{erfc}}

	\begin{acronym}
\acro{5G-NR}{5G New Radio}
\acro{3GPP}{3rd Generation Partnership Project}
\acro{AC}{address coding}
\acro{ACF}{autocorrelation function}
\acro{ACR}{autocorrelation receiver}
\acro{ADC}{analog-to-digital converter}
\acrodef{aic}[AIC]{Analog-to-Information Converter}     
\acro{AIC}[AIC]{Akaike information criterion}
\acro{aric}[ARIC]{asymmetric restricted isometry constant}
\acro{arip}[ARIP]{asymmetric restricted isometry property}

\acro{ARQ}{automatic repeat request}
\acro{AUB}{asymptotic union bound}
\acrodef{awgn}[AWGN]{Additive White Gaussian Noise}     
\acro{AWGN}{additive white Gaussian noise}

\acro{IoT}{Internet of Things}
\acro{IoE}{Internet of Everything}
\acro{ZF}{Zero-Forcing}

\acro{mmW}{Millimeter Waves}
\acro{SQP}{sequential quadratic programming}

\acro{APSK}[PSK]{asymmetric PSK} 

\acro{waric}[AWRICs]{asymmetric weak restricted isometry constants}
\acro{warip}[AWRIP]{asymmetric weak restricted isometry property}
\acro{BCH}{Bose, Chaudhuri, and Hocquenghem}        
\acro{BCHC}[BCHSC]{BCH based source coding}
\acro{BEP}{bit error probability}
\acro{BFC}{block fading channel}
\acro{BG}[BG]{Bernoulli-Gaussian}
\acro{BGG}{Bernoulli-Generalized Gaussian}
\acro{BPAM}{binary pulse amplitude modulation}
\acro{BPDN}{Basis Pursuit Denoising}
\acro{BPPM}{binary pulse position modulation}
\acro{BPSK}{binary phase shift keying}
\acro{BPZF}{bandpass zonal filter}
\acro{BSC}{binary symmetric channels}              
\acro{BU}[BU]{Bernoulli-uniform}
\acro{BER}{bit error rate}
\acro{BS}{base station}
\acro{EI}{exposure index}

\acro{CP}{Cyclic Prefix}
\acrodef{cdf}[CDF]{cumulative distribution function}   
\acro{CDF}{cumulative distribution function}
\acrodef{c.d.f.}[CDF]{cumulative distribution function}
\acro{CCDF}{complementary cumulative distribution function}
\acrodef{ccdf}[CCDF]{complementary CDF}               
\acrodef{c.c.d.f.}[CCDF]{complementary cumulative distribution function}
\acro{CD}{cooperative diversity}

\acro{CDMA}{Code Division Multiple Access}
\acro{ch.f.}{characteristic function}
\acro{CIR}{channel impulse response}
\acro{cosamp}[CoSaMP]{compressive sampling matching pursuit}
\acro{CR}{cognitive radio}
\acro{cs}[CS]{compressed sensing}                   
\acrodef{cscapital}[CS]{Compressed sensing} 
\acrodef{CS}[CS]{compressed sensing}
\acro{CSI}{channel state information}
\acro{CCSDS}{consultative committee for space data systems}
\acro{CC}{convolutional coding}
\acro{Covid19}[COVID-19]{Coronavirus disease}

\acro{DAA}{detect and avoid}
\acro{DAB}{digital audio broadcasting}
\acro{DCT}{discrete cosine transform}
\acro{dft}[DFT]{discrete Fourier transform}
\acro{DR}{distortion-rate}
\acro{DS}{direct sequence}
\acro{DS-SS}{direct-sequence spread-spectrum}
\acro{DTR}{differential transmitted-reference}
\acro{DVB-H}{digital video broadcasting\,--\,handheld}
\acro{DVB-T}{digital video broadcasting\,--\,terrestrial}
\acro{DL}{downlink}
\acro{DSSS}{Direct Sequence Spread Spectrum}
\acro{DFT-s-OFDM}{Discrete Fourier Transform-spread-Orthogonal Frequency Division Multiplexing}
\acro{DAS}{distributed antenna system}
\acro{DNA}{Deoxyribonucleic Acid}

\acro{EC}{European Commission}
\acro{EED}[EED]{exact eigenvalues distribution}
\acro{EIRP}{Equivalent Isotropically Radiated Power}
\acro{ELP}{equivalent low-pass}
\acro{eMBB}{Enhanced Mobile Broadband}
\acro{EMF}{electric and magnetic fields}
\acro{EU}{European union}

\acro{FC}[FC]{fusion center}
\acro{FCC}{Federal Communications Commission}
\acro{FEC}{forward error correction}
\acro{FFT}{fast Fourier transform}
\acro{FH}{frequency-hopping}
\acro{FH-SS}{frequency-hopping spread-spectrum}
\acrodef{FS}{Frame synchronization}
\acro{FSsmall}[FS]{frame synchronization}  
\acro{FDMA}{Frequency Division Multiple Access}

\acro{GA}{Gaussian approximation}
\acro{GF}{Galois field }
\acro{GG}{Generalized-Gaussian}
\acro{GIC}[GIC]{generalized information criterion}
\acro{GLRT}{generalized likelihood ratio test}
\acro{GPS}{Global Positioning System}
\acro{GMSK}{Gaussian minimum shift keying}
\acro{GSMA}{Global System for Mobile communications Association}

\acro{HAP}{high altitude platform}

\acro{IDR}{information distortion-rate}
\acro{IFFT}{inverse fast Fourier transform}
\acro{iht}[IHT]{iterative hard thresholding}
\acro{i.i.d.}{independent, identically distributed}
\acro{IoT}{Internet of Things}                      
\acro{IR}{impulse radio}
\acro{lric}[LRIC]{lower restricted isometry constant}
\acro{lrict}[LRICt]{lower restricted isometry constant threshold}
\acro{ISI}{intersymbol interference}
\acro{ITU}{International Telecommunication Union}
\acro{ICNIRP}{International Commission on Non-Ionizing Radiation Protection}
\acro{IEEE}{Institute of Electrical and Electronics Engineers}
\acro{ICES}{IEEE international committee on electromagnetic safety}
\acro{IEC}{International Electrotechnical Commission}
\acro{IARC}{International Agency on Research on Cancer}
\acro{IS-95}{Interim Standard 95}

\acro{LEO}{low earth orbit}
\acro{LF}{likelihood function}
\acro{LLF}{log-likelihood function}
\acro{LLR}{log-likelihood ratio}
\acro{LLRT}{log-likelihood ratio test}
\acro{LOS}[LoS]{line-of-sight}
\acro{LRT}{likelihood ratio test}
\acro{wlric}[LWRIC]{lower weak restricted isometry constant}
\acro{wlrict}[LWRICt]{LWRIC threshold}
\acro{LPWAN}{low power wide area network}
\acro{LoRaWAN}{Low power long Range Wide Area Network}
\acro{NLOS}[NLoS]{non-line-of-sight}

\acro{MB}{multiband}
\acro{MC}{multicarrier}
\acro{MDS}{mixed distributed source}
\acro{MF}{matched filter}
\acro{m.g.f.}{moment generating function}
\acro{MI}{mutual information}
\acro{MIMO}{multiple-input multiple-output}
\acro{MISO}{multiple-input single-output}
\acrodef{maxs}[MJSO]{maximum joint support cardinality}                       
\acro{ML}[ML]{maximum likelihood}
\acro{MSE}{mean-square error}
\acro{MMSE}{minimum MSE}

\acro{MMV}{multiple measurement vectors}
\acrodef{MOS}{model order selection}
\acro{M-PSK}[${M}$-PSK]{$M$-ary phase shift keying}                       
\acro{M-APSK}[${M}$-PSK]{$M$-ary asymmetric PSK} 

\acro{M-QAM}[$M$-QAM]{$M$-ary quadrature amplitude modulation}
\acro{MRC}{maximal ratio combiner}                  
\acro{maxs}[MSO]{maximum sparsity order}                                      
\acro{M2M}{machine to machine}                                                
\acro{MUI}{multi-user interference}
\acro{mMTC}{massive Machine Type Communications}      
\acro{mm-Wave}{millimeter-wave}
\acro{MP}{mobile phone}
\acro{MPE}{maximum permissible exposure}
\acro{MAC}{media access control}
\acro{NB}{narrowband}
\acro{NBI}{narrowband interference}
\acro{NLA}{nonlinear sparse approximation}
\acro{NTIA}{National Telecommunications and Information Administration}
\acro{NTP}{National Toxicology Program}
\acro{NHS}{National Health Service}

\acro{OC}{optimum combining}                             
\acro{OC}{optimum combining}
\acro{ODE}{operational distortion-energy}
\acro{ODR}{operational distortion-rate}
\acro{OFDM}{orthogonal frequency-division multiplexing}
\acro{omp}[OMP]{orthogonal matching pursuit}
\acro{OSMP}[OSMP]{orthogonal subspace matching pursuit}
\acro{OQAM}{offset quadrature amplitude modulation}
\acro{OQPSK}{offset QPSK}
\acro{OFDMA}{Orthogonal Frequency-division Multiple Access}
\acro{OPEX}{Operating Expenditures}
\acro{OQPSK/PM}{OQPSK with phase modulation}

\acro{PAM}{pulse amplitude modulation}
\acro{PAR}{peak-to-average ratio}
\acrodef{pdf}[PDF]{probability density function}                      
\acro{PDF}{probability density function}
\acrodef{p.d.f.}[PDF]{probability distribution function}
\acro{PDP}{power dispersion profile}
\acro{PMF}{probability mass function}                             
\acrodef{p.m.f.}[PMF]{probability mass function}
\acro{PN}{pseudo-noise}
\acro{PPM}{pulse position modulation}
\acro{PRake}{Partial Rake}
\acro{PSD}{power spectral density}
\acro{PSEP}{pairwise synchronization error probability}
\acro{PSK}{phase shift keying}
\acro{PD}{power density}
\acro{8-PSK}[$8$-PSK]{$8$-phase shift keying}

\acro{FSK}{frequency shift keying}

\acro{QAM}{Quadrature Amplitude Modulation}
\acro{QPSK}{quadrature phase shift keying}
\acro{OQPSK/PM}{OQPSK with phase modulator }

\acro{RD}[RD]{raw data}
\acro{RDL}{"random data limit"}
\acro{ric}[RIC]{restricted isometry constant}
\acro{rict}[RICt]{restricted isometry constant threshold}
\acro{rip}[RIP]{restricted isometry property}
\acro{ROC}{receiver operating characteristic}
\acro{rq}[RQ]{Raleigh quotient}
\acro{RS}[RS]{Reed-Solomon}
\acro{RSC}[RSSC]{RS based source coding}
\acro{r.v.}{random variable}                               
\acro{R.V.}{random vector}
\acro{RMS}{root mean square}
\acro{RFR}{radiofrequency radiation}
\acro{RIS}{reconfigurable intelligent surface}
\acro{RNA}{RiboNucleic Acid}

\acro{SA}[SA-Music]{subspace-augmented MUSIC with OSMP}
\acro{SCBSES}[SCBSES]{Source Compression Based Syndrome Encoding Scheme}
\acro{SCM}{sample covariance matrix}
\acro{SEP}{symbol error probability}
\acro{SG}[SG]{sparse-land Gaussian model}
\acro{SIMO}{single-input multiple-output}
\acro{SINR}{signal-to-interference plus noise ratio}
\acro{SIR}{signal-to-interference ratio}
\acro{SISO}{single-input single-output}
\acro{SMV}{single measurement vector}
\acro{SNR}[\textrm{SNR}]{signal-to-noise ratio} 
\acro{sp}[SP]{subspace pursuit}
\acro{SS}{spread spectrum}
\acro{SW}{sync word}
\acro{SAR}{specific absorption rate}
\acro{SSB}{synchronization signal block}

\acro{TH}{time-hopping}
\acro{ToA}{time-of-arrival}
\acro{TR}{transmitted-reference}
\acro{TW}{Tracy-Widom}
\acro{TWDT}{TW Distribution Tail}
\acro{TCM}{trellis coded modulation}
\acro{TDD}{time-division duplexing}
\acro{TDMA}{Time Division Multiple Access}

\acro{UAV}{unmanned aerial vehicle}
\acro{uric}[URIC]{upper restricted isometry constant}
\acro{urict}[URICt]{upper restricted isometry constant threshold}
\acro{UWB}{ultrawide band}
\acro{UWBcap}[UWB]{Ultrawide band}   
\acro{URLLC}{Ultra Reliable Low Latency Communications}
         
\acro{wuric}[UWRIC]{upper weak restricted isometry constant}
\acro{wurict}[UWRICt]{UWRIC threshold}                
\acro{UE}{user equipment}
\acro{UL}{uplink}

\acro{WiM}[WiM]{weigh-in-motion}
\acro{WLAN}{wireless local area network}
\acro{wm}[WM]{Wishart matrix}                               
\acroplural{wm}[WM]{Wishart matrices}
\acro{WMAN}{wireless metropolitan area network}
\acro{WPAN}{wireless personal area network}
\acro{wric}[WRIC]{weak restricted isometry constant}
\acro{wrict}[WRICt]{weak restricted isometry constant thresholds}
\acro{wrip}[WRIP]{weak restricted isometry property}
\acro{WSN}{wireless sensor network}                        
\acro{WSS}{wide-sense stationary}
\acro{WHO}{World Health Organization}
\acro{Wi-Fi}{wireless fidelity}

\acro{sss}[SpaSoSEnc]{sparse source syndrome encoding}

\acro{VLC}{visible light communication}
\acro{VPN}{virtual private network} 
\acro{RF}{radio frequency}
\acro{FSO}{free space optics}
\acro{IoST}{Internet of space things}

\acro{GSM}{Global System for Mobile Communications}
\acro{2G}{second-generation cellular network}
\acro{3G}{third-generation cellular network}
\acro{4G}{fourth-generation cellular network}
\acro{5G}{5th-generation cellular networks}	
\acro{gNB}{next generation node B base station}
\acro{NR}{New Radio}
\acro{UMTS}{Universal Mobile Telecommunications Service}
\acro{LTE}{Long Term Evolution}

\acro{KKT}{Karush–Kuhn–Tucker}
\acro{UAV}{Unmanned Aerial Vehicle}

\acro{IPM}{interior-point method}
\acro{QoS}{quality of service}
\acro{ZF}{zero-forcing}
\end{acronym}

	\title{EMF-Aware Cellular Networks in RIS-Assisted Environments}
	
	\author{Hussam Ibraiwish, \IEEEmembership{Student {Member},~IEEE} Ahmed~Elzanaty,~\IEEEmembership{Member,~IEEE}, Yazan H. Al-Badarneh,~\IEEEmembership{Member,~IEEE}, Mohamed-Slim Alouini,  \IEEEmembership{Fellow,~IEEE}

		\thanks{H. Ibraiwish, and M.-S. Alouini  are with the Computer, Electrical and Mathematical Sciences and Engineering (CEMSE) Division, King Abdullah University of Science and Technology (KAUST), Thuwal 23955, Saudi Arabia (e-mail: \{hussam.Ibraiwish, slim.alouini\}@kaust.edu.sa).
			
			A. Elzanaty is with the Institute for Communication Systems (ICS), University of Surrey, Guildford GU2 7XH, U.K. (e-mail: a.elzanaty@surrey.ac.uk). He was with CEMSE Division, KAUST. 
			
			Y. H. Al-Badarneh is with the Department of Electrical
			Engineering, The University of Jordan, Amman 11942, Jordan (e-mail: yalbadarneh@ju.edu.jo).}
		
		\thanks{The research has been funded by  Competitive Research Grant (CRG) 2020.} 
	}

	\markboth{Accepted for publications in the IEEE Communications Letters}{Ibraiwish}

	\maketitle
	
	\begin{abstract}
		The  deployment of the \ac{5G} and beyond has triggered health concerns due to the \ac{EMF} exposure. In this paper, we propose a novel architecture to minimize the population exposure to \ac{EMF} by considering a smart radio environment with a \ac{RIS}. Then, we  optimize  the \ac{RIS} phases to minimize the exposure in terms of the \ac{EI} while maintaining a minimum target quality of service. The proposed scheme achieves up to $20\%$ reduction in \ac{EI} compared to schemes without  \acp{RIS}.
		
	\end{abstract}
	
	\acresetall 
	
	\section{Introduction}	
	The \ac{IARC} classified \ac{RFR} as {\em possibly carcinogenic to humans (Group 2B)} based on experimental studies regarding non-thermal impacts \cite{IARC:11,NTP:18a}. Recently, the wide deployment of \ac{5G} triggers health concerns among the population regarding  \ac{EMF} exposure \cite{ElzanatyLucaSlim:20,elzanaty20215g}. Such fear increases for networks that adopt higher frequency bands such as  \acl{mm-Wave}s. In these bands, users have  to increase their transmit power to cope with the high path loss, escalating their \ac{EMF} exposure .

	In the literature, some schemes have been proposed to mitigate the \ac{EMF} exposure. For instance, the authors in \cite{SamImaHelImr:17} design a resource allocation scheme to minimize the \ac{EMF} exposure while maintaining a minimum \ac{QoS} for users. In \cite{5723738},  a beamforming based technique is proposed for minimizing the \ac{EMF} exposure. On the other hand, the authors in \cite{6932477} consider ferrite meta-material between the mobile phone and human head for reducing the exposure. Recently, unmanned aerial vehicles have been used to reduce the exposure \cite{lou2021green}.
	
	Most previous works consider legacy networks with only operational optimization such as resource allocation. However, in order to significantly minimize the exposure an architectural development is required, especially for beyond \ac{5G} networks operating at higher frequency bands. 
	
	Recently, architectural-based solutions exploiting what  are called  \acp{RIS} have been proposed to  enhance communication systems performance \cite{8796365}. Unlike natural surfaces where the incident angle  equals to  reflected angle, a \ac{RIS} can reflect the incident  wave toward a specified direction depending on the phase shift induced on its surface~\cite{najafi2020physics}.

	In this paper, we propose a novel architecture where a \ac{RIS} is exploited to minimize the \ac{EMF} exposure.
	we consider the \ac{UL} of a cellular network in a \ac{RIS} assisted environment involving single-antenna mobile users and a multi-antenna \ac{BS}. The contribution of this work can be summarized as follows.
	\begin{itemize}
		\item We propose a novel architecture that exploits a \ac{RIS} with optimized phase design  to minimize the overall exposure by passively focusing users' signals  toward the \ac{BS} to achieve the target data rate with minimal transmit power and exposure. 
		\item  We develop a dual gradient descent algorithm to optimize  the \ac{RIS} phases and derive closed-form expressions for the gradient, Hessian matrix, and optimized step size.
	\end{itemize}

	For notations, vectors and matrices are denoted by bold small and capital letters, respectively. The ${k}^{\text{th}}$ column and row of $\mathbf{X}$ are written as $\mathbf{x}_{k}$ and $\mathbf{x}_{(k)}$, respectively, while $x_{m,n}$ represents the element in the $m^{\text{th}}$ row and $n^{\text{th}}$ column. Similarly,   $x_{n}$ represents the ${n}^{\text{th}}$ element of a  vector $\mathbf{x}$. The first and second norm of vector $\mathbf{x}$ are represented as $\|\mathbf{x}\|_1$ and $\|\mathbf{x}\|$, respectively. The Frobenius norm,  pseudo inverse, and Hermitian transpose of $\mathbf{X}$ are represented as $\|\mathbf{X}\|_\mathrm{F}$, $\mathbf{X}^+$ , and $\mathbf{X}^{\mathrm{H}}$, respectively. The Kronecker product between $\mathbf{X}$ and $\mathbf{Y}$ is denoted by $\mathbf{X}\otimes\mathbf{Y}$. Finally, the real and imaginary part of a complex matrix $\mathbf{Z}$ are denoted by  $\mathbf{Z}^\mathrm{Re}$ and $\mathbf{Z}^\mathrm{Im}$, respectively.

	\section{Proposed RIS-assisted Architecture}
	\begin{figure}[t!]
		\centering
		\scalebox{.6}{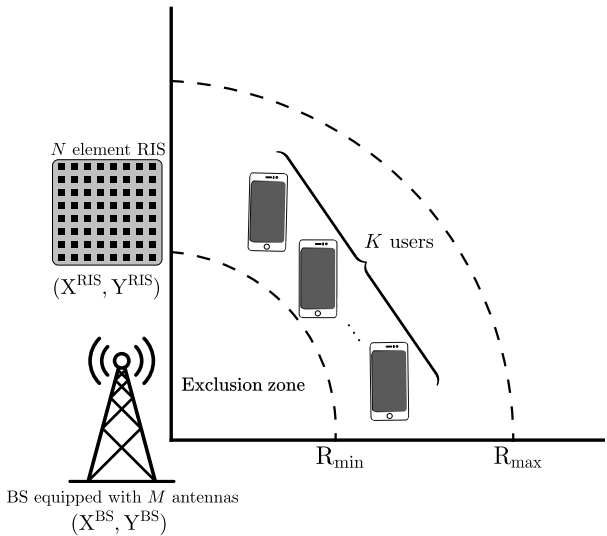}
		\caption{The system model of the proposed cellular network with a \ac{RIS}.}
		\label{scenario}
	\end{figure}
	We consider a single \ac{BS}, equipped with $M$ antennas, located at $(\mathrm{X^{BS},Y^{BS}})$ serving $K$ users each with a single antenna. A \ac{RIS} with $N$ reflective elements located at $(\mathrm{X^{RIS},Y^{RIS}})$ is considered to enhance the uplink channel, as in Fig.~\ref{scenario}. The received signal at the \ac{BS} can be written as
	\begin{equation}
		\mathbf{y}=\left(\Htwo\,\boldsymbol{\Phi}(\boldsymbol{\theta})\,\Hone+\Hthree \right)\, \x+ \n,
	\end{equation}
	where  $x_{k}$ represents the complex transmitted symbol from user $k$ and $\x \in \mathbb{C}^{K\times 1}$. The transmit power of user $k$ can be simply computed as $p_k = \mathbb{E}\{{x}^{\phantom{*}}_{k} {x}^{*}_{k} \}$. The matrices $\Hone \in \mathbb{C}^{N\times K}$, $\Htwo \in \mathbb{C}^{M\times N}$, and $\Hthree \in \mathbb{C}^{M\times K}$ represent the channel matrix between the users and \ac{RIS}, \ac{RIS} and \ac{BS}, and \ac{BS} and users, respectively.
	The element $\theta_n$ is the phase shift induced by the $n^{\text{th}}$ element of the \ac{RIS}, $\boldsymbol{\theta} \in \mathbb{R}^{N\times1}$, $\phi_{n,n}  \triangleq		e^{j\theta_n}$, and $\boldsymbol{\Phi}(\boldsymbol{\theta}) \in \mathbb{C}^{N\times N}$ is a diagonal matrix. The vector $\n \in \mathbb{C}^{M\times1}$ represents the \ac{AWGN} at the \ac{BS}, where all entries of $\n$ are complex Gaussian random variables with zero means and $\sigma^2$ variances. The \ac{BS} estimates users' signals by applying a linear estimator, i.e., multiplying   $\mathbf{y}$ by a beamforming matrix $\mathbf{G} \in \mathbb{C}^{K\times M}$.
	
	\section{EMF-Aware Design with RIS}
	A single user exposure can be quantified by the \ac{SAR}, which is defined as the power absorbed per mass of exposed tissue $(\mathrm{W/Kg})$. The SAR for a certain user can be modeled as a summation of the exposure caused by the cellular phone itself due to the \ac{UL}, i.e., $\mathrm{SAR^{UL}}$, and  the exposure resulting form the base station due to the \ac{DL}, i.e., $\mathrm{SAR^{DL}}$. The exposure for the $k^{
		\text{th}}$ user is
	\begin{equation}
		\mathrm{SAR}_k =\mathrm{SAR}_k^\mathrm{UL}+\mathrm{SAR}_k^\mathrm{DL}\triangleq\mathrm{SAR}_k^\mathrm{ref,UL} p_k+\mathrm{SAR}_k^\mathrm{ref,DL}\ s_k,
	\end{equation}
	where $p_{k}$ is the transmit power in $\mathrm{W}$, $\mathrm{SAR}_k^\mathrm{ref,UL}$ is the normalized $\mathrm{SAR}_k^\mathrm{UL}$ (i.e., the induced $\mathrm{SAR}_k^\mathrm{UL}$ when $p_k$ equals  $1\,\mathrm{W}$), $s_k$ is the received power density  from the \ac{BS} in $\mathrm{W/m^2}$, and  $\mathrm{SAR}_k^\mathrm{ref,DL}$ 
	is the normalized $\mathrm{SAR}_k^\mathrm{DL}$ (i.e., $\mathrm{SAR}_k^\mathrm{DL}$ when $s_k$ equals $1\,\mathrm{W/m^2}$). To account for the population exposure, we consider the \ac{EI} metric \cite{vermeerenlow}
	\begin{align}
		\EI&\triangleq \sum_{k=1}^{K}  \mathrm{SAR}_k  \  [\textrm{W/Kg}].
	\end{align}
	
	The \ac{BS} transmit power is significantly higher than that of users. However, the received power density $s_k$ can be quite low, because users are usually far from the \ac{BS}, i.e. located outside the exclusion zone. According to the Federal Communications Commission (FCC), the received power density near typical cellular towers is usually less than $10\, m\mathrm{W/m^2}$. Using a typical value of $\mathrm{SAR}_k^\mathrm{ref,DL}$ in \cite{vermeerenlow}, we have a $\mathrm{SAR}_k^\mathrm{DL}$ of $42\ \mu \mathrm{W/Kg}$.
	On the other hand, for the maximum transmit power for a mobile phone of $200\, m\mathrm{W}$,  we have $\mathrm{SAR}_k^\mathrm{UL}=1060\ \mu \mathrm{W/Kg}$, which is significantly higher than the \ac{DL} exposure, as also shown in \cite{lou2021green}. Hence, we consider only the \ac{UL} exposure. Now, the \ac{EI} can be written as\footnote{Since only the \ac{UL} exposure is considered, we remove the {UL} superscript.}
	\begin{equation}
		\EI\left(\mathbf{p}\right)= \sum_{k=1}^{K} \mathrm{SAR}^{\mathrm{ref}}_{k}\, p_{k}\quad [\textrm{W/Kg}],
	\end{equation}
	Then, the optimization problem can be formulated as
	\begin{subequations}
		\label{eq:originalopt}
		\begin{alignat}{2}
			&\underset{\mathbf{p},\boldsymbol{\theta}, \mathbf{G}}{\text{minimize}}        &\qquad& \EI\left(\mathbf{p}\right)\label{eq:optProb}\\
			&\textrm{subject to} &      &r_{k}\left(\boldsymbol{\theta}, \mathbf{G}, \mathbf{p}\right) \triangleq \log_{2} \left(1+\mathrm{SINR}_{k}\right) \geq   r^{\text{th}}_{k} \label{eq:constraint1}\\
			&               &      & p_{k} \leq p_{\max},  \forall k \in \mathbb{K} \triangleq \{1,2,\cdots,K\} \label{eq:constraint2},
		\end{alignat}
	\end{subequations}
	where $r_{k}$ in bits/Hz is the actual spectral efficiency for user $k$, $r^{\text{th}}_{k}$ is the minimum spectral efficiency required by user $k$, 
	$p_{\max}$ is the maximum allowable transmit power of users' devices, which also limits the exposure experienced by each user, and $\mathrm{SINR}_{k}$ is  the \ac{SINR} of the $k^{\text{th}}$ user. The  \ac{SINR} can be written as
	\begin{equation}\label{SINR}
		\mathrm{SINR}_{k}\!=\! \frac{{|\mathbf{g}_{(k)}(
				\Htwo\,\boldsymbol{\Phi}(\boldsymbol{\theta})\,\mathbf{h}^{\mathrm{u}}_k+\mathbf{h}^{\mathrm{d}}_k )\, x_k
				|}^2}{{{\sigma }^2\!\left\|{\mathbf{g}}_{(k)}\right\|}^2\!+\!\sum^K_{i=1,i\neq k}{{|\mathbf{g}_{(k)}(
					\Htwo\,\boldsymbol{\Phi}(\boldsymbol{\theta})\,\mathbf{h}^{\mathrm{u}}_i\!+\!\mathbf{h}^{\mathrm{d}}_i ) x_i
					|}^2}},
	\end{equation}
	where  $\mathbf{g}_{(k)}$ is the $k^{\text{th}}$ row of the beamforming matrix $\mathbf{G}$.
	
	\section{Optimization Algorithm}
	
	\subsection{Problem Reduction}
	
	The problem is non-convex and requires joint optimization of the beamforming, users' transmit power, and \ac{RIS} phases. We propose an alternate optimization algorithm where we first design the beamforming and user allocated power for a fixed \ac{RIS} phases profile. Then, we optimize the \ac{RIS} phases and iterate between these two processes till convergence. 
	
	For a fixed \ac{RIS} phase profile, the optimal beamforming vectors and allocated powers for this problem can be obtained as in \cite{4133009}. Nevertheless, this method  requires high computational complexity iterative algorithm. Also, it does not suppress the interference from various users at the \ac{BS}, further complicating the problem.
	In this regard, we  design the beamformer so that it eliminates the cross-interference from various users at the \ac{BS}, i.e.,  \ac{ZF} beamforming.
	Hence, the beamforming matrix at the \ac{BS} can be written as  \cite{ngo2015massive}
	\begin{equation}\label{beamformer}
		\mathbf{G}(\boldsymbol{\theta})={(\Htwo\, \boldsymbol{\Phi}(\boldsymbol{\theta})\, \Hone+\Hthree)}^{+} ,\quad \text{for\,} N \geq K\ \text{and\,}\  M \geq K.
	\end{equation}
	By noting that ${|\mathbf{g}_{(k)}(
		\Htwo\,\boldsymbol{\Phi}(\boldsymbol{\theta})\,\mathbf{h}^u_k+\mathbf{h}^d_k )\, x_k
		|}^2=p_k$ and $\sum^K_{i=1,i\neq k}{{|\mathbf{g}_{(k)}(
			\Htwo\,\boldsymbol{\Phi}(\boldsymbol{\theta})\,\mathbf{h}^{\mathrm{u}}_i\!+\!\mathbf{h}^{\mathrm{d}}_i ) x_i
			|}^2}=0$, the \ac{SINR} can be written as $\mathrm{SINR}_{k}={p_k}\,{{{\sigma }^{-2}\,\left\|{\mathbf{g}}_{(k)}\right\|}^{-2}}$.
	
	Since the objective function and  achieved data rate are monotonically increasing in the users' transmit power, the optimal $p_k$ is the power that achieves the minimum required data rate $r^{\text{th}}_{k}$. Otherwise, this will lead to an unnecessary exposure rise. Thus, the power of the $k^{\text{th}}$ user is set to
	\begin{align}
		p_k\left(\boldsymbol{\theta}\right) = (2^{r^{\text{th}}_{k}}-1)\,\sigma^2\,{\left\|{\mathbf{g}}_{(k)}\right\|}^2 \triangleq \widetilde{\sigma}_k^2\, {\left\|{\mathbf{g}}_{(k)}\right\|}^2, \,\forall\, k \in \mathbb{K}.
		\label{power}
	\end{align}
	Accordingly, we can eliminate the data rate constraint by substituting  \eqref{power} in \eqref{eq:optProb}. 
	After representing both $ \mathbf{G}$ and $\mathbf{p}$ as functions of $\boldsymbol{\theta}$ as in \eqref{beamformer} and \eqref{power}, the optimization problem is relaxed to
	\begin{subequations}
		\label{eq:relaxedopt}
		\begin{alignat}{2}
			&\underset{\boldsymbol{\theta}}{\text{minimize}}        &\qquad& 
			\sum_{k=1}^{K} \mathrm{SAR}^{\text{ref}}_{k}\  \widetilde{\sigma}_k^2\,{\left\|{\mathbf{g}}_{(k)}\right\|}^2\label{eq:final_optProb}\\
			&\textrm{subject to} &      &\widetilde{\sigma}_k^2\,{\left\|{\mathbf{g}}_{(k)}\right\|}^2\,\leq\,p_{\max}. \label{eq:final_constraint1}
		\end{alignat}
	\end{subequations}
	
	Although  problem \eqref{eq:relaxedopt} is not identically equivalent to \eqref{eq:originalopt}, it can lead to a solution that sufficiently reduces the \ac{EI} with low  computational complexity.
	
	As we can see in \eqref{power}, larger $ \widetilde{\sigma}_k^2\,$ values necessitate higher power to be transmitted from the mobile and possibly that power will exceed  $p_{\max}$. Hence, the maximum  $p_{k}$ is forced to be $p_{\max}$, which can result in a rate, $r_{k}$, that is less than the target rate, $r_{k}^\mathrm{th}$, i.e., the problem becomes infeasible. Therefore,  we define the rate satisfaction ratio for the users as $\sum_{k=1}^Kr_{k}/\sum_{k=1}^Kr_{k}^\mathrm{th}$, quantifying the solution feasibility. Even though the problem is still non-convex,  applying various convex optimization algorithms can significantly reduce the objective function, leading to a local minimum. The solution for the above problem must satisfies the  \ac{KKT} conditions where it must be a saddle point within the Lagrangian function.
	Thus, our new objective is to find a saddle point for the following Lagrangian function
	\begin{equation}
		{\mathcal{L}(\boldsymbol{\theta},\boldsymbol{\lambda})}=\ \EI\left(\mathbf{p}(\boldsymbol{\theta})\right)+\boldsymbol{\lambda}^\top(\mathbf{p}(\boldsymbol{\theta})-\mathbf{p}_{\max}), \label{lag}
	\end{equation}
	where $\boldsymbol{\lambda} \in {\mathbb{R}}^{K\times 1} $ represents the \ac{KKT} multipliers and $\mathbf{p}_{\max} \in \mathbb{R}^{K\times 1} $ is a vector with each entity being {$p_{\max}$}.
	
	In this work, we propose the dual gradient descent algorithm where we iteratively apply two phases \cite{dualgradient}. In the first phase, the Lagrangian function is minimized by applying the gradient descent considering only the optimization variables while in the second phase, the Lagrangian function is maximized by applying the gradient ascent considering only the multipliers, thus both gradient descent and ascent are adopted. At the $t^\mathrm{th}$ iteration, the algorithm updates the phases and multipliers as
	\begin{align}
		\boldsymbol{\theta}^{\{t\}}&=\boldsymbol{\theta}^{\{t-1\}}-\alpha \nabla_{\boldsymbol{\theta}}\,\mathcal{L}\left(\boldsymbol{\theta}^{\{t-1\}},\boldsymbol{\lambda}^{\{t-1\}}\right)\\
		\boldsymbol{\lambda}^{\{t\}}&=\boldsymbol{\lambda}^{\{t-1\}}+\beta \nabla_{\boldsymbol{\lambda}}\,\mathcal{L}\left(\boldsymbol{\theta}^{\{t\}},\boldsymbol{\lambda}^{\{t-1\}}\right),
	\end{align}
	where $\alpha$ and $\beta$ are the step sizes for updating the \ac{RIS} phases and multipliers, respectively. The gradient of the Lagrangian function with respect to both of the \ac{RIS} phases and  \ac{KKT} multipliers and the step sizes are investigated below.
	\subsection{Gradient of the Lagrangian Function}
	Let $a_k\triangleq  \left(\mathrm{SAR}^{\text{ref}}_{k}+ \lambda_k \right) \widetilde{\sigma}_k^2$,  $\mathbf{A} \triangleq \mathrm{diag}[a_1,a_2,...,a_k]$,  $\Honebar \triangleq \Hone \sqrt{\mathbf{\mathbf{A}}^{-1}}$, and $\Hthreebar \triangleq \Hthree \sqrt{\mathbf{\mathbf{A}}^{-1}}$, the Lagrangian function in \eqref{lag} can be simplified as
	\begin{equation}
		{\mathcal{L}\left(\boldsymbol{\theta},\boldsymbol{\lambda}\right)} 	=\mathrm{tr}\left(\mathbf{T}\right) \minus,
		\label{eq:lagrang}
	\end{equation}
	where  $\mathbf{T} \triangleq (\Q^H\Q )^{-1} $ and $ \Q \triangleq \Htwo\, \boldsymbol{\Phi}\, \Honebar+\Hthreebar$.  The derivative of the Lagrangian function with respect to $\theta_i$ is denoted by $\mathcal{L}'(i)$ and can be derived using matrix calculus as\footnote{For notational  simplicity, we drop the dependency on $\boldsymbol{\theta}$ and $\boldsymbol{\lambda}$ in $\mathcal{L}(\boldsymbol{\theta},\boldsymbol{\lambda})$.}
	\begin{equation}
		\mathcal{L}'(i)\!\triangleq\!\frac{\partial\, \mathcal{L}}{\partial\,\theta_i}\!=
		\mathfrak{R}\left\{\!\mathrm{tr} 
		\left(\!je^{j\theta _n} \left(  \mathbf{{h}}^{\mathrm{r}}_{i} \otimes \overline{\mathbf{h}}^{\mathrm{u}}_{(i)} \right) 
		\left({ \frac{\partial\,  \mathcal{L}} {\partial\,\Q} }\right) ^H \!\right)  \!\right\}, \label{gradient}
	\end{equation}
	where 
	\begin{equation}
		\frac{\partial\,  \mathcal{L}} {\partial\,\Q}=-2\, \Q\,{\mathbf{T}}^{2}, \label{dl/dq}
	\end{equation}
	as shown in the appendix. On the other hand, the derivative with respect to the multipliers can be written as
	\begin{equation}
		\frac{\partial\,\mathcal{L}}{\partial\,\lambda_i}=\widetilde{\sigma}_k^2\, {\left\|{\mathbf{g}}_{(i)}\right\|}^2 -p_{\max},\quad \forall i\in \{1,2,\cdots,N\}.
		\label{gradient_mul}
	\end{equation}
	\begin{algorithm}[t!]
		\DontPrintSemicolon
		\SetKwInOut{Input}{Input}\SetKwInOut{Output}{Output}
		\Input{$\boldsymbol{\theta}$, $\boldsymbol{\lambda}$}
		
		$\alpha^*\gets 0$, $\mathrm{EI_{min}} \gets \infty$ \algorithmiccomment{{\small \color{gray!30!black}Initialization}}\;
		
		$\mathbf{g}\gets \nabla_{\boldsymbol{\theta}}\mathcal{L}\left(\boldsymbol{\theta},\boldsymbol{\lambda}\right)$  \algorithmiccomment{{\small \color{gray!30!black}Compute the gradient at $\boldsymbol{\theta}$}}\;

		$\bar{\boldsymbol{\alpha}}\gets [0,
		{0.5\pi}/{ \mathrm{max}(|\mathbf{g}|)},
		{\pi}/{ \mathrm{max}(|\mathbf{g}|)}
		]$ \algorithmiccomment{{\small \color{gray!30!black}Initialize the guessing points of $\alpha^*$. The points \! ${0.5\pi}/{ \mathrm{max}(|\mathbf{g}|)}$, ${\pi}/{ \mathrm{max}(|\mathbf{g}|)}$ were chosen so that the maximum updated phase will be shifted by $\pi/2$ and $\pi$, respectively.}}\;
		
		\For{$i \  	\mathrm{in}\   (1,\mathrm{length}(\bar{\boldsymbol{\alpha}}))			$} { 
			
			$\widetilde{\boldsymbol{\theta}} \gets \boldsymbol{\theta}-\bar{\alpha}_i\, \mathbf{g}$  \algorithmiccomment{{\small \color{gray!30!black}Compute $\widetilde{\boldsymbol{\theta}} $ using $\bar{\alpha}_i$ }} \;
			
			\If	{$\mathrm{EI}(\mathbf{p}(\widetilde{\boldsymbol{\theta}}))<\mathrm{EI}_{min}$}
			{
				$\widetilde{\mathbf{g}}\gets \nabla_{\boldsymbol{\theta}}\mathcal{L}(\widetilde{\boldsymbol{\theta}},\boldsymbol{\lambda})$ \algorithmiccomment{{\small \color{gray!30!black}Compute the gradient at $\widetilde{\boldsymbol{\theta}}$}}\;
				$\widetilde{\mathbf{H}}\gets \mathbf{H}_{\boldsymbol{\theta}}(\widetilde{\boldsymbol{\theta}},\boldsymbol{\lambda}) $\algorithmiccomment{{\small \color{gray!30!black}Compute the Hessian at $\widetilde{\boldsymbol{\theta}}$}}\;
				\If{$\mathbf{g}^\top\widetilde{\mathbf{H}}\,\mathbf{g}>0		$}
				{				
					$\alpha^* \gets  \bar{\alpha}_i+
					(\mathbf{g}^\top \widetilde{\mathbf{g}})/
					\mathbf{g}^\top\widetilde{\mathbf{H}}\,\mathbf{g}$\algorithmiccomment{{\small \color{gray!30!black}Compute $\alpha^*$}}\;
					
					$ \mathrm{EI_{min}} \gets \mathrm{EI}(\mathbf{p}(\widetilde{\boldsymbol{\theta}})) $ \algorithmiccomment{{\small \color{gray!30!black}update $\EI_\mathrm{min}$}}\;
				}
			}	
		}
		
		\Return $\alpha^*$
		\caption{Optimal Step Size Sub-Algorithm}
		\label{optiaml_step}
	\end{algorithm}
	\begin{algorithm}[t!]
		\DontPrintSemicolon
		\SetKwInOut{Input}{Input}\SetKwInOut{Output}{Output}
		\Input{$\Hone$, $\Htwo$, $\Hthree$, $\sigma$, $\gamma$, ${\mathrm{\mathrm{SAR}}^{\text{ref}}}_k$, $r_\mathrm{th} ,\ \forall k \in \mathbb{K}$ 		}
		
		$\boldsymbol{\theta} \gets \mathbf{0} \ ,\ \boldsymbol{\lambda} \gets \mathbf{0} $ \algorithmiccomment{{\small \color{gray!30!black}Initialize $\boldsymbol{\lambda}$ and $\boldsymbol{\theta}$}}\;
		
		$\beta^*=\gamma\,\sum_{k=1}^K {\mathrm{SAR}^{\text{ref}}_k}/K$ \algorithmiccomment{{\small \color{gray!30!black}Compute 	$\beta^*$}}\;
		
		\While{$\mathrm{stopping\ criterion\ not\ met}$}
		{
			Compute $\mathcal{L}'_i,\,\forall i\in(1,2,..., N)$ using \eqref{gradient}\;
			Compute  $\alpha^*$ using algorithm \eqref{optiaml_step}\;
			
			$\boldsymbol{\theta}\leftarrow\boldsymbol{\theta}-\alpha^* \nabla_{\boldsymbol{\theta}}\mathcal{L}\left(\boldsymbol{\theta},\boldsymbol{\lambda}\right)$  	\algorithmiccomment{{\small \color{gray!30!black}update $\boldsymbol{\theta}$}}\;
			
			Compute ${\partial\, \mathcal{L}}/{\partial\,\lambda_k},\,\forall k\in\mathbb{K}$ using  \eqref{gradient_mul}\;
			
			$\boldsymbol{\lambda}\leftarrow\mathrm{max}(\mathbf{0}\,,\, \boldsymbol{\lambda}+\beta^* \nabla_{\boldsymbol{\lambda}}\mathcal{L}\left(\boldsymbol{\theta},\boldsymbol{\lambda}\right))$ \algorithmiccomment{{\small \color{gray!30!black}update $\boldsymbol{\lambda}$}}\;

			Compute $p_k,\,\forall k \in \mathbb{K} $ using \eqref{power}\;
			
			\If {$p_k<p_\mathrm{max}$}{
				$\lambda_k \gets 0 $  \algorithmiccomment{{\small \color{gray!30!black}$\forall k \in \mathbb{K}$}}	
		}}
		
		$\mathbf{p}\leftarrow\mathrm{min}(\mathbf{p},\mathbf{p}_{\mathrm{max}})$\;

		\Return ${\boldsymbol{\theta}}$, $\mathbf{p}$
		\caption{EMF Reduction Algorithm }
	\end{algorithm}
	\subsection{Optimal Step Size}
	Regarding the step size for the  phase update,    we approximate the Lagrangian function around an initial step size $\bar{\alpha}$ by a quadratic function. Then, the optimal step size $\alpha^*$ of the approximated Lagrangian function can be found as 
	\begin{equation}
		\alpha^*=
		\bar{\alpha}
		+\frac{(\nabla_{\boldsymbol{\theta}}\mathcal{L}(
			\boldsymbol{\theta},\boldsymbol{\lambda}))^\top\ 
			\nabla_{\boldsymbol{\theta}}\mathcal{L}(\widetilde{\boldsymbol{\theta}},\boldsymbol{\lambda})}      {(\nabla_{\boldsymbol{\theta}}\mathcal{L}(
			\boldsymbol{\theta},\boldsymbol{\lambda}))^\top\ \mathbf{H}_{\boldsymbol{\theta}}(\widetilde{\boldsymbol{\theta}},\boldsymbol{\lambda})\  \nabla_{\boldsymbol{\theta}}\mathcal{L}(
			\boldsymbol{\theta},\boldsymbol{\lambda}) } ,
		\label{eq:alphathetaopt}
	\end{equation}
	where $\widetilde{\boldsymbol{\theta}}= \boldsymbol{\theta}-\bar{\alpha} \nabla_{\boldsymbol{\theta}}\mathcal{L}\left(\boldsymbol{\theta},\boldsymbol{\lambda}\right)$ and $\mathbf{H}_{\boldsymbol{\theta}}(\widetilde{\boldsymbol{\theta}},\boldsymbol{\lambda})  \in \mathbb{R}^{N\times N}$  is the Hessian matrix of the Lagrangian function computed at $\widetilde{\boldsymbol{\theta}}$ and $\boldsymbol{\lambda}$. The elements of the Hessian matrix can be computed as 
	\begin{equation}
		\frac{\partial^2\mathcal{L}}{\partial\theta_v\partial\theta_u}
		\!=\!
		\begin{cases} 
			\mathfrak{R}\left\{ \psi_{v,u}  \right\} & , v\neq u \\
			\mathfrak{R}\left\{ \mathrm{tr}\left(j\mathbf{R}\mathbf{T}^2\Q^H\right)+ \psi_{v,v}  \right\}       & , v=u 
		\end{cases}
		,
	\end{equation}
	where
	\begin{align}
		\psi_{v,u} &\triangleq \mathrm{tr}\bigg(
		je^{j\theta_u} \left(  \mathbf{h}^{\mathrm{r}}_{u}\otimes \mathbf{\overline{\mathbf{h}}}^{\mathrm{u}}_{(u)} \right) 
		{\left({ \frac{\partial\, \mathcal{L}'(v)}  {\partial\,\Q} }\right)}^{H} \bigg),\\
		\frac{\mathcal{L}'(v)} {\partial\,\Q}&=
		\mathbf{RT}^2-\mathbf{QT}\left(
		\mathbf{TC}-\mathbf{CT}\right)\mathbf{T},
	\end{align}
	for   	$\mathbf{R}\triangleq -2je^{j\theta_v}  \left(  \mathbf{h}^{\mathrm{r}}_{v}\otimes \mathbf{\overline{h}}^{\mathrm{u}}_{(v)} \right) $ and 		${\mathbf{C} \triangleq	\mathbf{R}^H	\Q-\Q^H	\mathbf{R}}$.

	The  optimal step size in \eqref{eq:alphathetaopt} is computed considering a given initial step size. Algorithm~\ref{optiaml_step} identifies the procedure required to obtain an appropriate step size that minimizes the Lagrangian function. More precisely,  
	the proposed algorithm begins with three initial step sizes, then the one corresponding to the minimum \ac{EI} and positive curvature is chosen. In rare situations when all the initial guesses have a negative curvature, the phases will not be updated, i.e., $\alpha^*=0$, and the main algorithm will terminate.
	
	On the other hand,  reasonable values for the step size of the multipliers,  $\beta$, is on the order of the average reference \ac{SAR}. In particular, we set $\beta=\gamma\,\sum_{k=1}^K {\mathrm{SAR}^{\text{ref}}_k}/K$, where $\gamma$ is a small constant around unity  that determines how quickly the  \ac{KKT} multipliers should be updated. For small values of $\gamma$ (i.e., small multipliers), the Lagrangian will be closer to the main objective function, yielding smaller \ac{EI}, albeit with lower rate satisfaction ratio. On the other hand, larger $\gamma$ results in more feasible users. Thus, there is a trade off between the \ac{EI} and rate satisfaction ratio depending on the choice of $\gamma$. 
	\subsection{Complexity Analysis}
	The computational complexity of computing the gradient is $\mathcal{O}(NMK)$, while the computational complexity of computing the Hessian is $\mathcal{O}(N^2MK)$. Thus, our algorithm is overwhelmed by computing the Hessian matrix which is essential to compute the optimal step size. Therefore, the overall complexity of our algorithm is $\mathcal{O}(N^2MK)$. All the computations are performed by the \ac{BS}, where it starts by estimating $\Hone$, $\Htwo$ and $\Hthree$, then it optimizes the \ac{RIS} phases and feeds such information back to the \ac{RIS}.\footnote{For instance, the channels $\Hone$, $\Htwo$ and $\Hthree$ can be estimated using various methods as shown in \cite{wei2021channel} and the references therein.}
	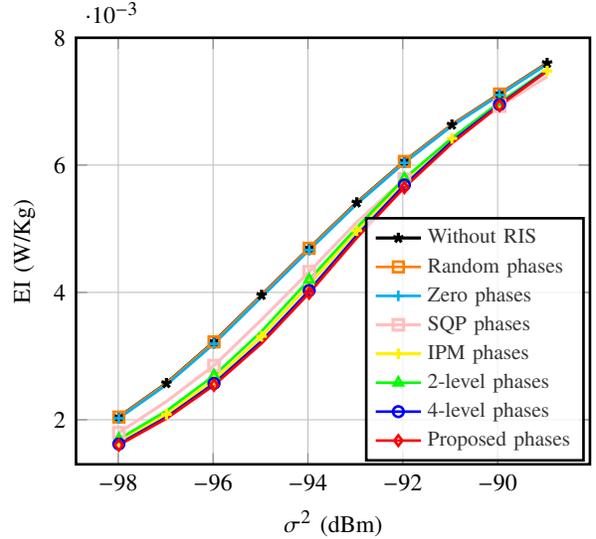
\begin{figure}
		\pgfplotsset{every axis/.append style={
		font=\small,
		line width=1pt,
	legend style={fill opacity=0.8, draw opacity=1,font=\footnotesize, at={(.99,0.58)}},legend cell align=left},
} 
\pgfplotsset{compat=1.13}
	\begin{tikzpicture}
\begin{axis}[
xlabel near ticks,
ylabel near ticks,
grid=major,
xlabel={$\sigma^2$ (dBm)},
ylabel={\ac{EI} (W/Kg)},
width=.95\linewidth,
legend entries={Without RIS
	,Random phases,
	Zero phases,
	SQP phases,
	IPM phases,
2-level phases,
4-level phases,
Proposed phases,
},
	ymin= .0013, ymax=0.008,
]
\addplot[black,mark=star,mark repeat=1,mark options=solid,very thick] table {Figures/FigPhasDesign/SAR_direct.dat};

	\addplot[orange,mark=square,mark repeat=2,mark options=solid] table {Figures/FigPhasDesign/SAR_mirror.dat};
	
	\addplot[cyan,mark=+,mark options=solid,mark repeat=2] table {Figures/FigPhasDesign/SAR_rand.dat};

\addplot[pink,mark=square,mark repeat=2,mark options=solid,very thick] table {Figures/FigPhasDesign/SAR_fix.dat};

\addplot[yellow,mark=+,mark repeat=1,mark options=solid,very thick] table {Figures/FigPhasDesign/SAR_fmincon.dat};

\addplot[green,mark=triangle,mark options=solid,mark repeat=2] table {Figures/FigPhasDesign/SAR_Q2.dat};

\addplot[blue,mark=o,mark options=solid,mark repeat=2] table {Figures/FigPhasDesign/SAR_Q4.dat};

\addplot[red,mark=diamond,mark repeat=2,mark options=solid] table {Figures/FigPhasDesign/SAR_optimized.dat};

\end{axis}
\end{tikzpicture}
		\caption{The \ac{EI} vs. $\sigma^2$ considering different scenarios, for $K=16$, $N=128$, and $M=32$.}
		\label{SARVsSNR}
	\end{figure}	
	\begin{figure}[t!]
		\pgfplotsset{every axis/.append style={
		font=\small,
		line width=1pt,
		legend style={font=\footnotesize, at={(0.45,0.6)}},legend cell align=left},
}
\pgfplotsset{compat=1.13}
	\begin{tikzpicture}
\begin{axis}[
xlabel near ticks,
ylabel near ticks,
grid=major,
xlabel={$\sigma^2$ (dBm)},
ylabel={Rate satisfaction ratio},
width=.95\linewidth,
legend entries={
	Proposed phases,
	4-level phases,
	2-level phases,
	IPM phases,
	SQP phases,
	Zero phases,
	Random phases,
	Without RIS
},
]

\addplot[red,mark=diamond,mark options=solid,mark repeat=1] table {Figures/FigPhasDesign/rate_optimized.dat};

\addplot[blue,mark=o,mark options=solid,mark repeat=1] table {Figures/FigPhasDesign/rate_Q4.dat};

\addplot[green,mark=triangle,mark options=solid,mark repeat=2] table {Figures/FigPhasDesign/rate_Q2.dat};

\addplot[yellow,mark=star,mark repeat=2,mark options=solid,very thick] table {Figures/FigPhasDesign/rate_fmincon.dat};
\addplot[pink,mark=square,mark repeat=2,mark options=solid,very thick] table {Figures/FigPhasDesign/rate_fix.dat};

\addplot[cyan,mark=+,mark repeat=1,mark options=solid] table {Figures/FigPhasDesign/rate_mirror.dat};
\addplot[orange,mark=square,mark repeat=2,mark options=solid] table {Figures/FigPhasDesign/rate_rand.dat};
\addplot[black,mark=star,mark repeat=2,mark options=solid,very thick] table {Figures/FigPhasDesign/rate_direct.dat};

\end{axis}
\end{tikzpicture}
		\caption{Rate satisfaction ratio vs. $\sigma^2$ considering different scenarios, for $K=16$, $N=128$, and $M=32$.}
		\label{satisfied data rate}
	\end{figure}
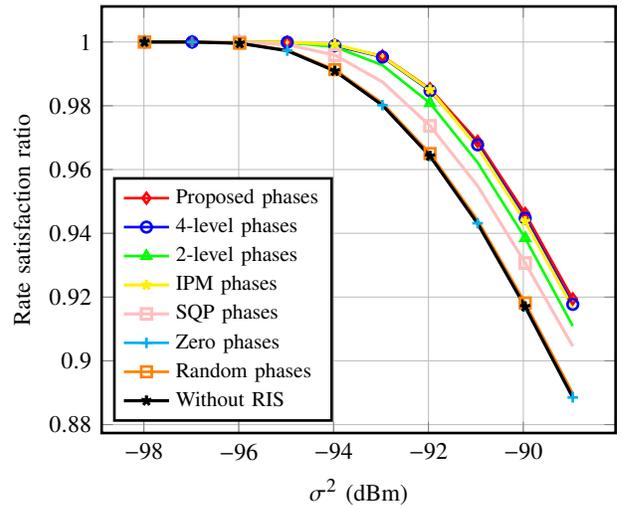
	\section{Numerical Results}
	In this section, we investigate the performance of our algorithm.
	We set the environment as shown in Fig.~\ref{scenario}, where there are $K=16$ active users uniformly distributed in the shown area. We set   $\mathrm{R_{min}}$ and $\mathrm{R_{max}}$ to $10$~$\mathrm{m}$ and $150$~$\mathrm{m}$, respectively. The \ac{BS} and \ac{RIS} are at a height of $10$~$\mathrm{m}$ and located at  $(0,0)$ and $(30,20)$~$\mathrm{m}$, respectively. We consider a Rician fading for the \ac{RIS}-users and \ac{RIS}-\ac{BS} links, as the paths involving \ac{RIS} usually have a \ac{LOS} component. Rayleigh fading is considered  
	for user-\ac{BS} links. In particular, we have
	\begin{align}
		\Htwo&=\mathrm{PL_{LOS}}(d^\mathrm{RB})\left(\sqrt{\frac{\kappa}{\kappa+1}}\,\mathbf{a}_N^\top\,\mathbf{a}_M+\sqrt{\frac{1}{\kappa+1}}\,     \tilde{\Htwo}   \right),\\
		{\mathbf{h}^{\mathrm{u}}_k}&=\mathrm{PL_{LOS}}(d^\mathrm{UR}_k)\left(\sqrt{\frac{\kappa}{\kappa+1}}\, \mathbf{a}_{k}+\sqrt{\frac{1}{\kappa+1}}\,     \tilde{{\mathbf{h}^{\mathrm{u}}_k}}   \right),\\
		{\mathbf{h}^{\mathrm{d}}_k}&=\mathrm{PL_{NLOS}}(d^{\mathrm{U_kB}})  \,\tilde{{\mathbf{h}^{\mathrm{u}}_k}},
	\end{align}where ${\mathbf{h}^{\mathrm{u}}_k}$ and ${\mathbf{h}^{\mathrm{d}}_k}$ are the $k^{\text{th}}$ columns of the matrices $\Hone$ and $\Hthree$, respectively, $d^\mathrm{RB}$ is the distance between the \ac{RIS} and the \ac{BS}, $d^\mathrm{UR}_k$  is the distance between the $k^{\text{th}}$ user and \ac{RIS}, $d^\mathrm{UB}_k$ is the distance between the $k^{\text{th}}$ user and the \ac{BS}, and $\kappa$ is the Rician factor which is set to $10$ in all  simulations. The functions $\mathrm{PL_{LOS}}(d)$ and $\mathrm{PL_{NLOS}}(d)$ are the path losses at distance $d$ for \ac{LOS} and  \ac{NLOS} links, respectively. Both $\mathrm{PL_{LOS}}(d)$ and $\mathrm{PL_{NLOS}}(d)$ are set according to the 3GPP model \cite{3GPP} as $
	\mathrm{PL_{LOS}}(d)=-35.6 + 22\log_{10}(d)$(dB) and $
	\mathrm{PL_{NLOS}}(d)=32.6 +36.7\log_{10}(d)$(dB). All the elements of $\tilde{\Htwo}$, $ \tilde{{\mathbf{h}^{\mathrm{u}}_k}} $ and $\tilde{{\mathbf{h}^{\mathrm{u}}_k}}$ are \ac{i.i.d.} complex Gaussian random variables with zero mean and unit variance.  Finally, $\mathbf{a}_M$, $\mathbf{a}_N$ and $\mathbf{a}_k$ are the  steering vectors \cite[(2)]{PARKER2017277}.
	
	In these simulations, each user is generated with a $75\%$  probability of being a data user  and a $25\%$ probability of being a voice user. Each data user has a minimum data rate of $600$ Mb/s and an allocated bandwidth of $100$ MHz. Each voice user has a minimum data rate of $13.3$ Kb/s and an allocated bandwidth of $7$ KHz. The reference \ac{SAR} can be found from  \cite[Table 27]{vermeerenlow} by averaging over different postures and population ages. The data users are assumed to use a mobile from a small distance in front of their torso, while voice users are assumed to put the mobile near their head resulting in $\SAR^{\mathrm{ref}}= 41\times10^{-4}$ and  $\SAR^{\mathrm{ref}}= 63\times10^{-4}$ $\mathrm{W/Kg}$ per unit power, respectively.
	\subsection{Effect of \ac{RIS} Phase Design}
	We compare the performance of our proposed algorithm in terms of the \ac{EI} and satisfied rate ratio with schemes without  \ac{RIS} and with various \ac{RIS} phase profiles, i.e.,  \textit{(i)} zero phases with $\boldsymbol{\theta}=\mathbf{0}$; \textit{(ii)} random phases, where the phase shifts are uniformly distributed from $0$ to $2\,\pi$;  \textit{(iii)} numerically optimized phases through the \ac{IPM} and \ac{SQP}. \textit{(iv)} $L$-level phases where the optimized \ac{RIS} phases are  discretized with $L$ levels uniform quantizer.
	\begin{figure}[t!]
		\pgfplotsset{every axis/.append style={
		font=\small,
		line width=1pt,
		legend style={font=\footnotesize, at={(0.98,0.98)}},legend cell align=left},
}
\pgfplotsset{compat=1.13}
	\begin{tikzpicture}
\begin{axis}[
ylabel style = {font=\small},
xlabel style = {font=\small},
xlabel near ticks,
ylabel near ticks,
grid=major,
xlabel={$N$},
ylabel={\ac{EI} (W/kg)},
width=.95\linewidth,
legend entries={M=16
,M=32
,M=64},
]
\addplot[blue,mark=star,mark repeat=2,mark options=solid,very thick] table {Figures/RIS_element_simulation/SARM16.dat};

\addplot[red,mark=square,mark repeat=2,mark options=solid] table {Figures/RIS_element_simulation/SARM32.dat};
	
\addplot[black,mark=o,mark repeat=2,mark options=solid] table {Figures/RIS_element_simulation/SARM64.dat};

\end{axis}
\end{tikzpicture}
		\caption{The \ac{EI} vs. the number of \ac{RIS} elements for different number of \ac{BS} antennas with $K=16$.}
		\label{SAR_elements}
	\end{figure}
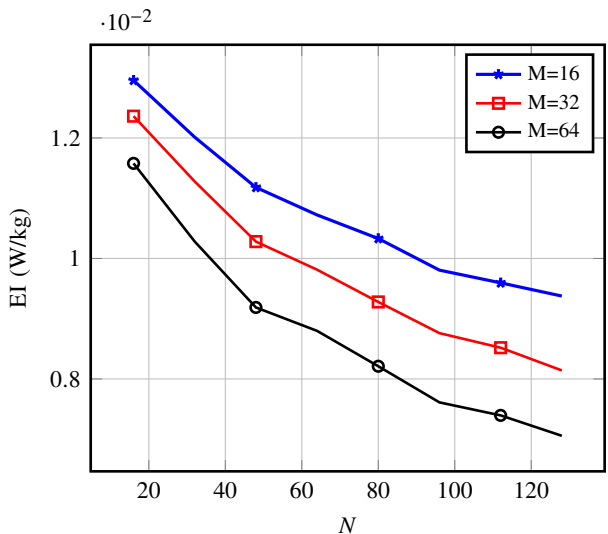
	\begin{figure}
		\pgfplotsset{every axis/.append style={
		font=\small,
		line width=1pt,
		legend style={font=\footnotesize, at={(0.98,0.26)}},legend cell align=left},
} 
\pgfplotsset{compat=1.13}
	\begin{tikzpicture}
\begin{axis}[
xlabel near ticks,
ylabel near ticks,
grid=major,
xlabel={$N$},
ylabel={Rate satisfaction ratio},
width=.95\linewidth,
legend entries={M=64
,M=32
,M=16},
]
\addplot[black,mark=o,mark repeat=2,mark options=solid] table {Figures/RIS_element_simulation/rateM64.dat};

\addplot[red,mark=square,mark repeat=2,mark options=solid] table {Figures/RIS_element_simulation/rateM32.dat};

\addplot[blue,mark=star,mark repeat=2,mark options=solid,very thick] table {Figures/RIS_element_simulation/rateM16.dat};

\end{axis}
\end{tikzpicture}	
		\caption{Rate satisfaction ratio vs. the number of \ac{RIS} elements for different number of \ac{BS} antennas with $K=16$.}
		\label{rate_elementss}
	\end{figure}
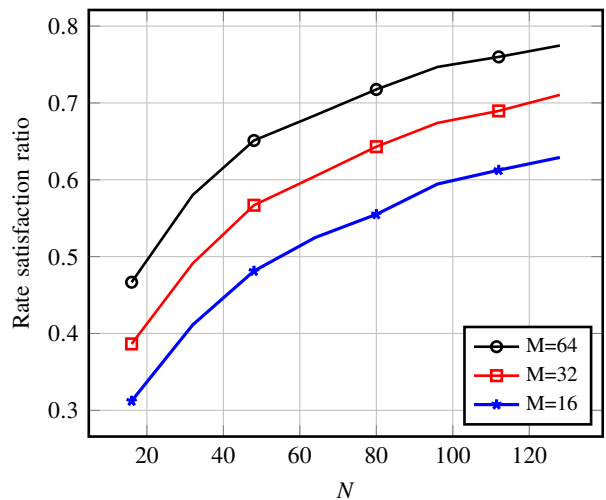
	In Fig.\ref{SARVsSNR}, the \ac{EI} for various noise levels is shown for different configurations. We notice that exploiting the \ac{RIS} with optimized phases reduces the \ac{EI} by $20\%$ compared to the schemes without \ac{RIS} and with non-optimized phases. Also, the \ac{EI} with random phases coincides with zero phases, because  the \ac{EI} is averaged  over a large number of channel realizations. Finally, increasing the number of  quantization levels yields results closer to the optimal solution. The data rate satisfaction ratio is shown in Fig. \ref{satisfied data rate}. The optimized phases design results in higher rate satisfaction ratio, and consequently higher achieved data rate,  compared to other schemes. 
	
	\subsection{Effect of Number of \ac{RIS} Elements }
	Here, we see how the number of \ac{RIS} elements $N$ affects the \ac{EI} and  rate satisfaction ratio for different number of \ac{BS} antennas, as shown in Fig.  \ref{SAR_elements} and Fig. \ref{rate_elementss}, respectively. We can see that increasing $N$ or $M$ always yields a better performance in terms of the \ac{EI} and the rate satisfaction ratio.
	\subsection{Algorithm Convergence }
\begin{figure}[t!]
		\pgfplotsset{every axis/.append style={
		font=\small,
		line width=1pt,
		legend style={font=\footnotesize, at={(0.98,0.98)}},legend cell align=left},
} 
\pgfplotsset{compat=1.13}
	\begin{tikzpicture}
\begin{axis}[
xlabel near ticks,
ylabel near ticks,
grid=major,
xlabel={Iteration number},
ylabel={\ac{EI} (W/kg)},
width=.95\linewidth,
legend entries={$\sigma^2=3.5 \times 10^{-13}$,$\sigma^2=3.18 \times 10^{-13}$,$\sigma^2=2.9 \times 10^{-13}$
},
]
\addplot[black,mark=diamond,mark repeat=5,mark options=solid] table {Figures/convergance/objective3.dat};

\addplot[blue,mark=square,mark repeat=5,mark options=solid] table {Figures/convergance/objective2.dat};

\addplot[red,mark=star,mark repeat=5,mark options=solid] table {Figures/convergance/objective1.dat};

\end{axis}
\end{tikzpicture}
		\caption{The \ac{EI} vs. iteration number, for $K=16$, $N=128$, and $M=32$.}
		\label{convergence_SAR}
	\end{figure}
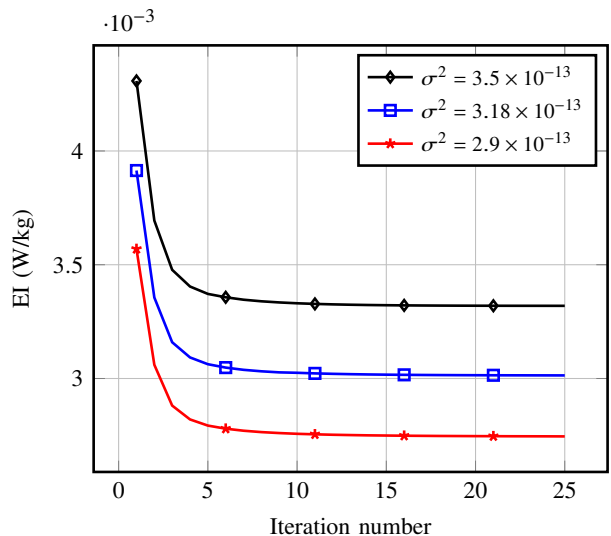

	The convergence of the dual gradient descent algorithm to a global minimum for convex optimization problems is discussed in \cite{dualgradient}. Although the problem is non convex the algorithm converges to a local minimum, as shown in Fig. \ref{convergence_SAR} for different noise power values.
	\section{Conclusion}
	We utilized a \ac{RIS} to minimize the users' exposure to \ac{EMF} for beyond \ac{5G} networks. We proposed a novel algorithm for minimizing the population \ac{UL} exposure in terms of the \ac{EI} while maintaining a target data rate. For instance, the proposed algorithm achieved a $20\%$  reduction in \ac{EI}  compared to the schemes without \ac{RIS} and with non-optimized phases. Furthermore, the optimized phase design achieved higher \ac{UL} data rates, even at harsh channel conditions.
		
\appendix
In order to compute the derivative of the Lagrangian with respect to the RIS phases, we reduce the Lagrangian as follows
\begin{align}
	{\mathcal{L}\left(\boldsymbol{\theta},\boldsymbol{\lambda}\right)}
	\!&\!= \sum_{k=1}^{K} \mathrm{SAR}^{\text{ref}}_{k}\, \widetilde{\sigma}_k^2{\left\|{\mathbf{g}}_{(k)}\right\|}^2\nonumber\!+\!  \sum_{k=1}^{K} \lambda_k\left( \widetilde{\sigma}_k^2{\left\|{\mathbf{g}}_{(k)}\right\|}^2 \!-\!p_{\max}\right)\nonumber\\
	& = \sum_{k=1}^{K} a_k{\left\|{\mathbf{g}}_{(k)}\right\|}^2 -p_{\max} \|\boldsymbol{\lambda}\|_1 \nonumber\\
	&=\!\mathrm{tr}(	\mathbf{G}^H\mathbf{A}	\mathbf{G} 	)\!-\!p_{\max} \|\boldsymbol{\lambda}\|_1\nonumber\\
	&=\!\mathrm{tr}(\mathbf{Q}^{+H}\mathbf{Q}^+)\!-\!p_{\max} \|\boldsymbol{\lambda}\|_1
	\nonumber\\
	&=\mathrm{tr}\left( \Q (\Q^H\Q )^{-1}  (\Q^H\Q )^{-1} \Q^H\right) \minus\nonumber\\
	&=\mathrm{tr}\left((\Q^H\Q )^{-1}\right) \minus,
	\label{eq:simplag}
\end{align}
which is equivalent  to \eqref{eq:lagrang}. Then we apply the chain rule as
\begin{align}
	\!\frac{\partial\,\mathcal{L}}{\partial\,\theta_n}\!&=\!\sum^M_{m=1}\!\sum^K_{k=1}\!\bigg(\!
	\frac{\partial\,\mathcal{L}}{\partial\,{\q^{\mathrm{Re}}}}
	\frac{\partial\,{\q^\mathrm{Re}}}{\partial\,\phin^\mathrm{Re}}
	\frac{\partial\,\phin^{\mathrm{Re}}}{\partial\,\theta_n}
	\!+\!\frac{\partial\,\mathcal{L}}{\partial\,{\q^{\mathrm{Re}}}}
	\frac{\partial\,{\q^{\mathrm{Re}}}}{\partial\,\phin^{\mathrm{Im}}}
	\frac{\partial\,\phin^{\mathrm{Im}}}{\partial\,\theta_n}
	\nonumber\\&+
	\frac{\partial\,\mathcal{L}}{\partial\,{\q^{\mathrm{Im}}}}
	\frac{\partial\,{\q^{\mathrm{Im}}}}{\partial\,\phin^{\mathrm{Re}}}
	\frac{\partial\,\phin^{\mathrm{Re}}}{\partial\,\theta_n}
	+
	\frac{\partial\,\mathcal{L}}{\partial\,{\q^{\mathrm{Im}}}}
	\frac{\partial\,{\q^{\mathrm{Im}}}}{\partial\,\phin^{\mathrm{Im}}}
	\frac{\partial\,\phin^{\mathrm{Im}}}{\partial\,\theta_n}\bigg).
	\label{chain}
\end{align}
Considering that $
\q=\sum^N_{i=1}h^\mathrm{r}_{m,n}\phi_{i,i} \overline{h}^\mathrm{u}_{n,k}$ we obtain
\begin{align}
	\frac{\partial\,{\q^{\mathrm{Re}}}}{\partial\,\phin^{\mathrm{Re}}}=&\,\frac{\partial\,{\q^{\mathrm{Im}}}}{\partial\,\phin^{\mathrm{Im}}}\quad=
	\mathfrak{Re} \left\{h^\mathrm{r}_{m,n}   \overline{h}^\mathrm{u}_{n,k}      \right\}
	,\\
	\frac{\partial\,{\q^{\mathrm{Im}}}}{\partial\,\phin^{\mathrm{Re}}}=&-	\frac{\partial\,{\q^{\mathrm{Re}}}}{\partial\,\phin^{\mathrm{Im}}}=
	\mathfrak{Im} \left\{ h^\mathrm{r}_{m,n}  \overline{h}^\mathrm{u}_{n,k}      \right\}
	,\\
	\frac{\partial\,\phin^{\mathrm{Re}}}{\partial\,\theta_n}=&-\mathrm{sin}(\theta_n)\ \ \ ,\ \ 	\frac{\partial\,\phin^{\mathrm{Im}}}{\partial\,\theta_n}=\mathrm{cos}(\theta_n).
\end{align}
To find  $\partial\,\mathcal{L}/\partial\,\q^\mathrm{Re}$ and $\partial\,\mathcal{L}/\partial\,\q^\mathrm{Im}$, we define the following 
\begin{align}
	\frac{\partial\,\mathcal{L}}{\partial\,{\q}} &\triangleq 	\frac{\partial\,\mathcal{L}}{\partial\,\q^{\mathrm{Re}}} +j
	\frac{\partial\,\mathcal{L}}{\partial\,\q^{\mathrm{Im}} } \\
	\frac{\partial\,\mathcal{L}}{\partial\,\Q} &\triangleq 
	\begin{pmatrix}
		\frac{\partial\,\mathcal{L}}{\partial\,{q_{1,1}}} & 	\frac{\partial\,\mathcal{L}}{\partial\,{q_{1,2}}} &\cdots &	\frac{\partial\,\mathcal{L}}{\partial\,{q_{1,K}}}
		\\
		\frac{\partial\,\mathcal{L}}{\partial\,{q_{2,1}}}& 	\frac{\partial\,\mathcal{L}}{\partial\,{q_{2,2}}} &\cdots &	\frac{\partial\,\mathcal{L}}{\partial\,{q_{2,K}}}
		\\
		\vdots & \vdots &\ddots&\vdots\\
		\frac{\partial\,\mathcal{L}}{\partial\,{q_{M,1}}}& 	\frac{\partial\,\mathcal{L}}{\partial\,{q_{M,2}}} &\cdots &	\frac{\partial\,\mathcal{L}}{\partial\,{q_{M,K}}}
	\end{pmatrix}.
\end{align}
For ${\partial\,\mathcal{L}}/{\partial\,\Q}\,$,  \eqref{gradient} can be obtained directly through matrix calculus \cite{IMM2012-03274}. After substituting in \eqref{chain}, we get 
\begin{equation}
	\frac{\partial\, \mathcal{L}}{\partial\,\theta_i}=
	\left\{
	\sum^M_{m=1}\sum^K_{k=1} j e^{j\theta_n} h^\mathrm{r}_{m,n}\overline{h}^\mathrm{u}_{n,k}
	\left(\frac{\partial\mathcal{L}}{\partial \,q_{m,k}}\right)^*
	\right\},
\end{equation}
which can be further reduced to \eqref{gradient}.

\bibliographystyle{IEEEtran}
\enlargethispage{0cm}
\bibliography{IEEEabrv,emf_ref.bib,Elzanaty_bibliography.bib}
\end{document}